# Area per Lipid in DPPC-Cholesterol Bilayers: Analytical Approach

B.B. Kheyfets, S.I. Mukhin[1]

Theoretical Physics and Quantum Technologies Department, NUST "MISIS", 119049 Moscow, Russia

Area per molecule in a DPPC-Cholesterol bilayers depends non-linearly on the cholesterol concentration. Using flexible strings model of lipid membranes we calculate area per molecule in DPPC-Cholesterol mixtures in the biologically relevant concentrations range. Few parameters of the model are optimized for a perfect agreement with the area per lipid data available from molecular dynamics simulations. Lateral pressure at the hydrophilic interface, $\gamma$, is taken to be proportional to the cholesterol concentration. Non-linearity arises as a consequence of the non-linear dependence of thermodynamical equilibrium area of molecules on $\gamma$. DPPC lipid is modeled as flexible string of finite thickness and a given bending rigidity, cholesterol molecule is modeled as rigid rod with finite thickness and infinite rigidity. Using parameters fitted to reproduce area per molecule dependence on cholesterol concentration, we had further calculated our model predictions for the NMR order parameter of DPPC lipid chains and coefficient of thermal area expansion. The microscopic nature of the model allows to consider a broad range of thermodynamic phenomena.

## INTRODUCTION

The lateral lipid membrane organization of multi-component lipid membranes that contain cholesterol is subjected to thorough investigation because of its significance for understanding of different vital membrane properties. Cholesterol is present in plasma membranes of higher eukaryotes [1–3]. The concentration of cholesterol largely varies between membranes of different cells and tissues [4]. Depending on the exact lipid composition, the plasma membranes of higher eukaryotes may contain 20-50 mol% of cholesterol [5].

Cholesterol plays a major role in the regulation of the fluidity and mechanical stiffness of membranes [3,6,7]. The ability of cholesterol to influence the fluidity of membranes is thought to be important in many biological processes including cell fusion [8], development of Alzheimer's disease [9,10], activity of the sodium pump [11], phase separation leading to raft formation [2,12], the functioning of the raft-embedded proteins [13] to name just a few.

When cholesterol is added to phospholipid membrane, the area per lipid decreases (egg-lecithin [14], DMPC [15], DPPC [16]). The area per lipid dependence on cholesterol concentration might be complex for some lipids and lipid mixtures but is monotonous for DPPC [17]. The main effects observed are a significant ordering of the DPPC chains [18], a reduced fraction of gauche bonds [16], a reduced surface area per lipid [19].

In DPPC membrane at low cholesterol concentrations (10-15 mol%) long-range lateral order is disrupted. Then at 18 mol%, a continuous cohesive liquid-ordered phase is formed. Finally, at 40 mol% cholesterol the membrane condenses to a gel phase [20].

---

[1] e-mail: i.m.sergei.m@gmail.com

The chemical structures of cholesterol and DPPC are shown on Fig. 1. Optimal membrane location of cholesterol in a membrane is determined by the desolvation free energy [21]. Cholesterol spans approximately one leaflet of the membrane, with its hydroxyl group protruding into the polar region of the bilayer [22].

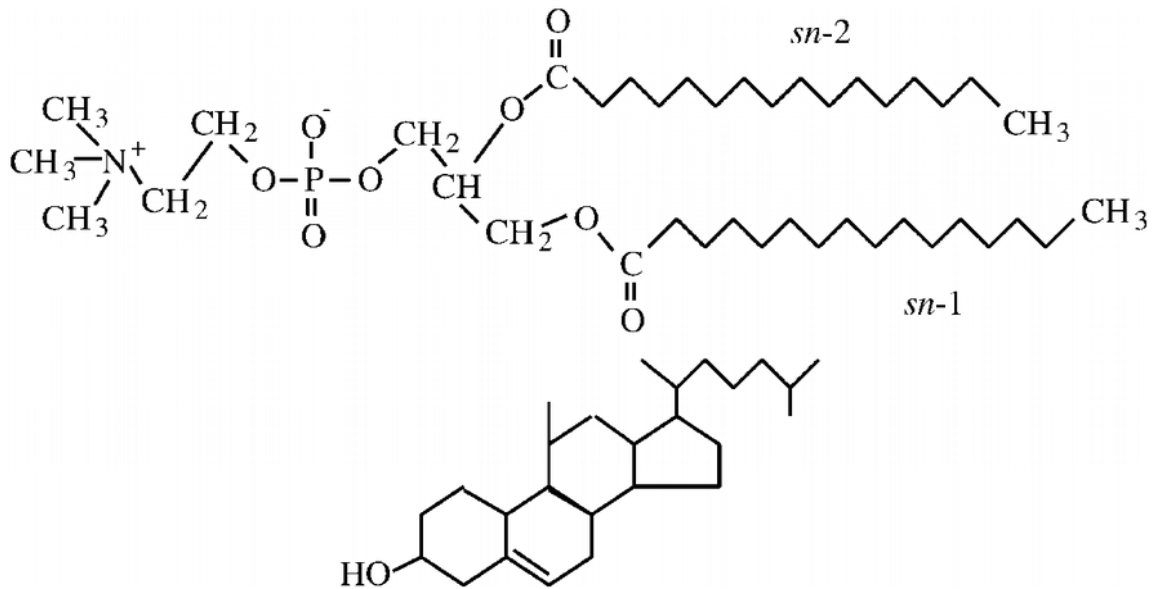

Fig. 1: Chemical structures of DPPC (upper) and cholesterol (lower).

The area per lipid dependence on concentration can be mimicked semi-quantitatively with regular solution theory using a condensed complexes model [17]. To the best of our knowledge a microscopic theory describing cholesterol condensing effect on DPPC membrane in a full range of biologically relevant cholesterol concentrations does not exist.

In this paper we use microscopic flexible strings model [23,24] to calculate analytically area per lipid in the DPPC-cholesterol mixtures in the broad concentration range of cholesterol. The theory does not allow for a description of the rich phase behavior exhibited by the DPPC-cholesterol mixtures: such as transition from liquid-disordered state to a liquid-ordered phase with high cholesterol concentration. However, it does capture the condensing effect. More over, if a few parameters are tailored to the molecular dynamics data on area per lipid [25] the agreement with the data is astonishingly good. The theory being microscopic then might be used to calculate a thermodynamical quantity of interest, such as lipid chains order parameter, lateral pressure profile, coefficient of thermal area expansion [23], pressure needed to make a pore [26], bending modulus [27], etc.

## FLEXIBLE STRINGS MODEL

Flexible strings model is a microscopic model of lipid in a membrane. In this section we give a survey of the model relevant for present paper. For more details the reader is referred to the original publications [23,24].

One can model phospholipid in a membrane as a flexible string of some incompressible core area, $A_n$, and some bending rigidity $K_f$. Subjecting the string to boundary conditions allows for a calculation of partition function, $Z$, of the string's oscillations parallel to the membrane's surface. The string on the average sweeps a surface area, $A$ (see Fig. 2). Requiring that pressure of the chain oscillations, $P$, on the neighboring lipids is balanced by the attraction of lipid heads, $\gamma$, one can calculate the area of the lipid (see Fig. 3).

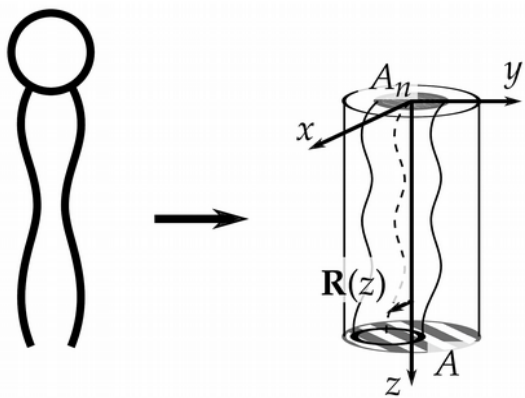

Fig. 2: Hydrocarbon tail as a flexible string.

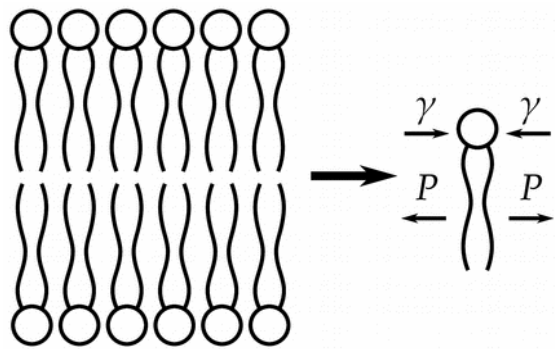

Fig. 3: In equilibrium attraction of the heads is balanced by a repulsion of the tails: $P = \gamma$.

Unsaturated hydrocarbon chains are not modeled directly: since the whole lipid is modeled as a single string (see Fig. 2). Instead, they influence the values of two constants: incompressible area of the lipid, $A_n$, and the bending rigidity $K_f$. All information about polar group of lipid is contained in a single parameter, $\gamma$ (see Fig. 3), which is lateral pressure at the hydrophilic interface.

The chain is described in terms of the deviation of the string's centers from a vertical, $\mathbf{R}(z)$ (see Fig. 2). Energy functional of the string is written as a sum of kinetic energy, bending energy [28] and interactions with neighboring chains which is modeled as a mean-field quadratic potential:

$$E_t = \int_0^L \left[ \frac{\rho \dot{\mathbf{R}}^2(z)}{2} + \frac{K_f}{2} \left( \frac{\partial^2 \mathbf{R}}{\partial z^2} \right)^2 + \frac{B\mathbf{R}^2}{2} \right] dz \qquad (1)$$

Here $L$ is width of the hydrocarbon part of the monolayer; $\rho$ is linear density of the hydrocarbon chain; $K_f$ is bending rigidity of the chain [24]; $B$ is a parameter of interaction lipid with neighboring chains to be determined self-consistently. The quadratic form for this interaction is used in polymer theory [29]. Nor peristaltic modes (thickness fluctuations), nor interlayer friction are taken into account. The integration is not performed for the $x$ and $y$ axis, because corresponded deviations of the centers of the string are considered small, and this assumption holds within the theory.

Using boundary conditions for a string it is possible to re-write potential energy in Eq. 1 in operator form. Interpreting its eigenfunctions as elementary oscillation modes, one can calculate the partition function, $Z$. The partition function might also be written as a path integral over all chain conformations. Considering $\frac{\partial F_t}{\partial B}$ using the two expressions for partition functions leads to the equation which relates area per lipid with a parameter of interlipid interaction, $B$:

$$b = \frac{1}{4v^{4/3}(\sqrt{a} - 1)^{8/3}} \qquad (2)$$

Where the following dimensionless variables were introduced:

$$a = \frac{A}{A_n}, \quad b = B\frac{L^4}{K_f}, \quad v = \frac{K_f A_n}{\pi k_B T L^3} \qquad (3)$$

Free energy of the string is a sum of free energy of oscillations and surface energy:

$$F_T = F_t + \gamma A \qquad (4)$$

here $\gamma$ is a microscopic surface tension. On the basis of undulation analysis it is estimated to be 50 dyn/cm [30] for DPPC at 323 K. Requiring free energy to be a minimum at equilibrium with respect to area of lipid one arrives at simple condition

$$\gamma = P \equiv -\frac{\partial F_t}{\partial A} \equiv \frac{k_B T}{A_n} \frac{1}{3v^{1/3}\sqrt{a}(\sqrt{a} - 1)^{5/3}} \qquad (5)$$

solving which an area per lipid might be finally obtained. The two main input parameters are incompressible area, $A_n$, and bending rigidity, $K_f$.

Cholesterol might be modeled as a string with infinite rigidity $K_f = \infty$: rigid rod. In this case conformations with bending possess an infinite energy and thus are eliminated. Hence, one is left with just two terms in the energy functional [26]:

$$E_t = \int_0^L \left[ \frac{\rho \dot{\mathbf{R}}^2(z)}{2} + \frac{B\mathbf{R}^2}{2} \right] dz \qquad (6)$$

The overall formalism holds the same. Eq. 2 becomes

$$B = \frac{\pi k_B T}{L A_n (\sqrt{a} - 1)^2} \quad (7)$$

(since for rigid rods $K_f = \infty$, one can't use dimensionless b as defined in Eq. 3). This expression might also be obtained from energy functional with finite bending rigidity Eq. 1: large $K_f$ corresponds to the opposite limit for $\nu$: $\nu \gg 1$, as compared with the limit $\nu \ll 1$ used in derivation of Eq. 2. Since $K_f = \infty$ then $\nu \gg 1$ (see Eq. 3) which leads exactly to the Eq. 7.

Rigid rod analogue of Eq. 5 is

$$\gamma = P \equiv -\frac{\partial F_t}{\partial A} \equiv \frac{k_B T}{A_n} \frac{1}{a - \sqrt{a}} \quad (8)$$

Rigid rod approximation, Eq. 8, allows one to find area of the lipid analytically:

$$a = \frac{1}{4}\left(1 + \sqrt{1 + 4\frac{k_B T}{A_n \gamma}}\right)^2 \quad (9)$$

Despite the rod is rigid mean area per lipid does not equal to the incompressible area, since the rod might move in lateral direction as a whole.

## CALCULATION

Consider a bilayer membrane consisting of DPPC and cholesterol. Let us denote $\gamma_D^0$ as surface tension at hydrophilic tension at a pure DPPC membrane.

At large concentration cholesterol doesn't form membranes and is known to be in a condensed state. Nevertheless, we can make an analytical extrapolation and consider pure cholesterol "membrane". Let us denote its surface tension at hydrophilic tension as $\gamma_C^0$.

Since $\gamma$ is a macroscopic parameter membrane composed of cholesterol and DPPC will have some effective $\gamma$ which might be measured using e.g. molecular dynamics simulations [30]. Assuming it to be proportional to the component's concentrations we write:

$$\gamma(c) = \gamma_C^0 c + \gamma_D^0 [1 - c] \quad (10)$$

where $c$ is cholesterol concentration.

Area per molecule is by definition

$$A(c) = A_C(c) c + A_D(c)[1 - c] \quad (11)$$

where $A_i(c)$ is found by solving

$$\frac{\partial F_t}{\partial A} + \gamma(c) = 0 \quad (12)$$

here $\frac{\partial F_t}{\partial A}$ is taken from Eq. 5 for DPPC lipid or Eq. 8 for cholesterol molecule.

Fig. 4 shows the result of solving Eq. 11 numerically.

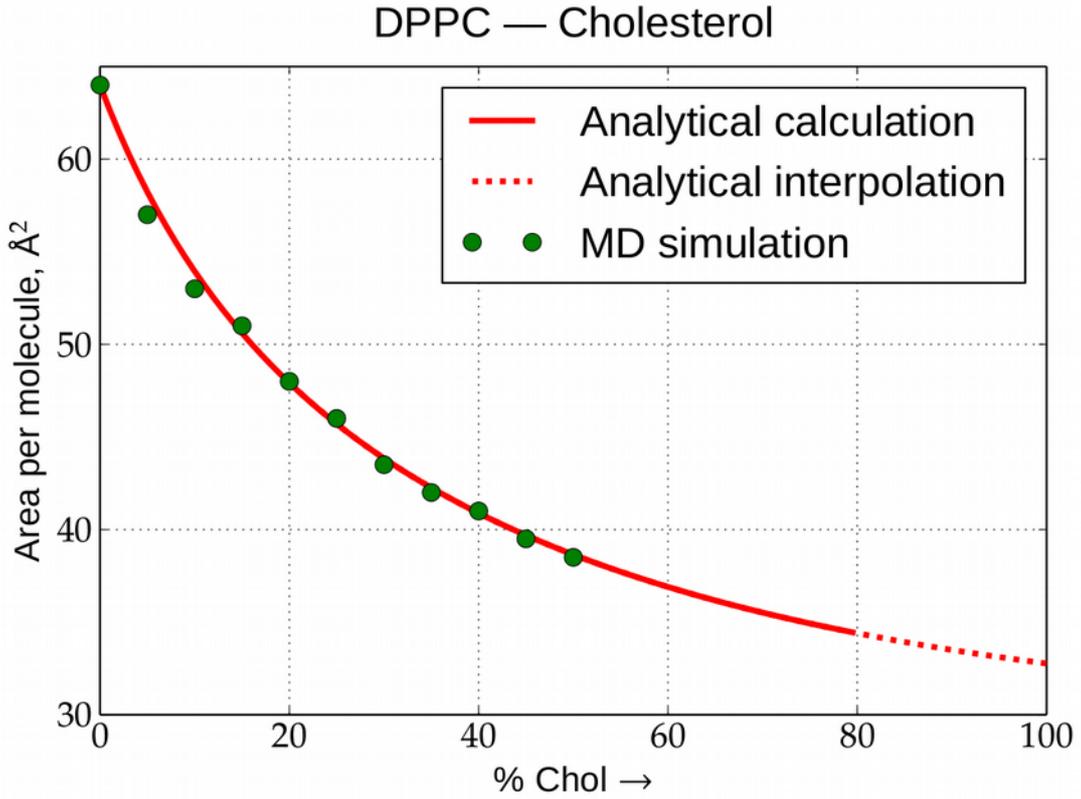

Fig. 4: Solid line: calculated area per molecule from Eq. 11 for biologically relevant cholesterol concentration range. Dotted line: analytical extrapolation of Eq. 11 into experimentally unattainable cholesterol-rich range. Green circles: averaged results of molecular dynamics simulations [25]. The parameters we used are: $A_{nC} = 28$ Å$^2$, $A_{nD} = 20$ Å$^2$, $\gamma_C^0 = 180$ erg/cm$^2$, $\gamma_D^0 = 32$ erg/cm$^2$, $K_f^{\text{DPPC}} = k_B T L/3$, $T = 323$ K, $L=17.2$ Å.

In order to find out the underlying physics of the numerical result on Fig. 4 we search for a condition which makes a calculated curve go below the ideal solution curve — the straight curve that connects area per molecule at pure DPPC and pure cholesterol membranes.

The equation for an ideal curve is

$$A_{\text{ideal}}(c) = (A_C^0 - A_D^0)c + A_D^0 \qquad (13)$$

we want to find a condition which guarantees that

$$\frac{\partial A(0)}{\partial c} < \frac{\partial A_{\text{ideal}}}{\partial c} \qquad (14)$$

(here $A(c)$ is defined in Eq. 11).

Taking derivative with respect to cholesterol concentration, $c$, of both sides of Eq. 11 and substituting $c = 0$ gives:

$$\frac{\partial A(0)}{\partial c} \equiv A_C(0) - A_D(0) + \frac{\partial A_D(0)}{\partial c} < A_C^0 - A_D^0 \qquad (15)$$

Partial derivative of $A_D$ with respect to $c$ might be calculated by differentiating equilibrium condition Eq. 12:

$$\frac{\partial^2 F_D(0)}{\partial A^2}\frac{\partial A_D(0)}{\partial c} + \frac{\partial \gamma}{\partial c} = 0 \tag{16}$$

Here $\frac{\partial \gamma}{\partial c} = \gamma_C^0 - \gamma_D^0$ (see Eq. 10) and we get an expression for $\frac{\partial A_D}{\partial c}$. Substituting it into inequality Eq. 15 gives

$$A_C(0) - A_C^0 - \frac{\gamma_C^0 - \gamma_D^0}{\frac{\partial^2 F_D(0)}{\partial A^2}} < 0 \tag{17}$$

where we took into account that $A_D^0$ is same as $A_D(0)$. Here

$$\frac{\partial^2 F_D}{\partial A^2} = \frac{k_B T}{A_n^2}\frac{4}{9v^{1/3}a(\sqrt{a}-1)^{8/3}} \tag{18}$$

(assuming $a$ is relatively large).

The inequality Eq. 17 might be true or false depending on the parameters of the two strings: one which models DPPC lipid, and the other which models cholesterol molecule.

Since Eq. 5 is analytically insolvable with respect to $a$, one can't express $\frac{\partial^2 F_D}{\partial A^2}$ in terms of input parameters of the model. Hence, in order to get further insight one has to resort to the approximation: we consider a liquid-disordered limit, $a_D \equiv A_D/A_{nD} \gg 1$, for DPPC lipids.

Although for rigid rods, which model cholesterol molecules, there's an analytical expression for area per molecule, Eq. 9, this expression is too complicated to deal with. Hence, we consider a liquid-ordered limit for cholesterol: $a_C \equiv A_C/A_{nC} \approx 1$.

Let us mention that for DPPC the approximation $a_D \equiv A_D/A_{nD} \gg 1$ is actually bad, since experimentally it is: $a_D \approx 3.20$, while approximation $a_C \equiv A_C/A_{nC} \approx 1$ for cholesterol is more justified, since experimentally it is: $a_C \approx 1.17$. Still, considering a liquid-disordered limit for DPPC lipids, and liquid-ordered limit for cholesterol should capture the main difference between the two.

## Liquid-disordered limit

Using Eq. 5 one can introduce a dimensionless parameter:

$$g_D = 3v^{1/3}\frac{A_{nD}\gamma_D^0}{k_B T} \equiv \frac{1}{\sqrt{a_D}(\sqrt{a_D}-1)^{5/3}} \tag{19}$$

Raising this equation to the $-3/5$ degree gives:

$$\frac{1}{g^{3/5}} = a_D^{4/5} - a_D^{3/10} \tag{20}$$

Assuming a liquid-disordered limit, $a_D \equiv A_D/A_{nD} \gg 1$, we write:

$$a_D \approx g^{-3/4} \tag{21}$$

Note that since $a_D \gg 1$ then $g \ll 1$. Using these conditions together with Eq. 21 we can re-write Eq. 18:

$$\frac{\partial^2 F_D}{\partial A^2} = \frac{k_B T}{A_{nD}^2} \frac{4 g_D^{7/4}}{9 v^{1/3}} \qquad (22)$$

## Liquid-ordered limit

Looking at Eq. 9 and implying $a_C \approx 1$ one finds that

$$g_C(\gamma) \equiv \frac{A_{nC}\gamma}{k_B T} \gg 1 \qquad (23)$$

Expanding Eq. 9 in $g_C$ then leads to

$$a_C(\gamma) = 1 + \frac{2}{g_C} \qquad (24)$$

## Condition for derivative

Substituting Eq. 22 and 24 into inequality Eq. 17 finally leads to

$$\frac{1}{\gamma_C^0} < \frac{3^{1/4}}{8} \frac{A_{nD}^{1/4}}{v^{1/4}(k_B T)^{1/4} \gamma_D^{0\,3/4}} \qquad (25)$$

This is a non-linearity condition: parameters of the two strings, one models DPPC lipid and the other models cholesterol molecule, have to satisfy it, for the area per lipid dependence on the cholesterol concentration to be below the ideal solution curve.

Despite this condition is derived using liquid-disordered and liquid-ordered limits, the parameters used to calculate area per molecule (see Fig. 4) satisfy it.

## RESULTS AND DISCUSSION

We consider bilayers of DPPC lipids and cholesterol molecules. We model DPPC lipid with flexible string and cholesterol molecule with rigid string, which supposed to model the chemical structure of the two molecules. We make an analytical extrapolation to model a pure cholesterol "membrane".

In order to describe interaction between the two lipids, which is evident from the area per molecule in the mixtures of DPPC lipids and cholesterol molecules, we assign different pressures at the hydrophilic interface for the pure membranes: $\gamma_C^0 > \gamma_D^0$. The quantity $\gamma$ is the only input parameter of flexible strings model which describes the influence of membrane on the lipid state.

We assumed a linear dependence of non-pure membrane on the cholesterol concentration (see Eq. 10). Area per molecule also depends linearly on the cholesterol concentration (see Eq. 11). Non-linearity in area per molecule of DPPC-cholesterol mixtures appears due to non-linear dependence of area per molecule on membrane's pressure at the hydrophilic interface for each of the component (see Eq. 5 and 8).

Few parameters of the model are optimized for a perfect agreement with data from molecular dynamics simulation [25] — see Fig. 4.

We also found a relation between parameters of the string, which leads to the non-linear area per molecule dependence on cholesterol concentrations (see Eq. 25). Namely, decreasing rigidity ($K_f$ through $\nu$, see Eq. 3) or pressure at the hydrophilic interface ($\gamma_D^0$) increases the effect of non-linear area per molecule dependence on cholesterol concentration; while increasing incompressible area of DPPC lipid — decreases this effect.

## Order parameter

In molecular dynamics study [18] it has been reported that merely increasing the tension in the DPPC membrane can't reproduce the order parameter of cholesterol-rich DPPC membrane. In this section we calculate NMR order parameter of the 60% DPPC, 40% cholesterol membrane.

There are at least two kinds of lipid order parameters in common use: one is molecular order parameter, $S_{mol}$, and the other is NMR order parameter, $S_{CD}$. The basic formula for the two is the same:

$$S = \frac{1}{2}\left(3\langle\cos^2\theta\rangle - 1\right) \tag{26}$$

but the angle is different.

In molecular order parameter the angle is defined as a deviation of the lipid chain from the averaged director of the lipid.

NMR order parameter measures the mobility of the deuterium, C-D, bond. The angle is in Eq. 26 is the angle between a given C-D bond and the normal to the membrane. The deuterium atom might be on either of the two C-H bonds of the given C atom in a polymer chain, that leads to the NMR order parameter at a given C atom is the mean order parameter of the two possible deuterium positions. If one would take into account that lipid is also rotating about its director and that trace of order parameter tensor is unity by definition, then one arrives at the so-called recursion relation [31]

$$S_{CD} = -\frac{1}{2}S_{mol} \tag{27}$$

Expression for $S_{mol}$ have been derived earlier [24]:

$$S_{mol} \approx 1 - 3k_BT\sum_n \frac{[R'_n(z)]^2}{E_n} \tag{28}$$

but in this paper we take into account tilt, which is reported to be 20° for 40% cholesterol DPPC membrane [18]. Denoting tilt with $\alpha$ we write:

$$\langle cos^2(\theta - \alpha)\rangle = [\cos\theta\cos\alpha - \sin\theta\sin\alpha]^2 \\ = \cos^2\alpha - \langle\sin^2\theta\rangle\cos 2\alpha \tag{29}$$

Assuming $\theta$ is small we substitute $\langle \sin^2 \theta \rangle$ with $\langle \tan^2 \theta \rangle$. Now using the result from the earlier work [24]:

$$\langle \tan^2 \theta \rangle = 2k_B T \sum_{n=0} \frac{[R'_n(z)]^2}{E_n} \qquad (30)$$

together with Eq. 27 we get

$$S_{CD} = -\frac{1}{4}\left(3\cos^2\alpha - 1 - 6\cos(2\alpha)k_B T \sum_{n=0} \frac{[R'_n(z)]^2}{E_n}\right) \qquad (31)$$

This result applies to flexible strings that we used to model DPPC lipids. The result is plotted in Fig. 5.

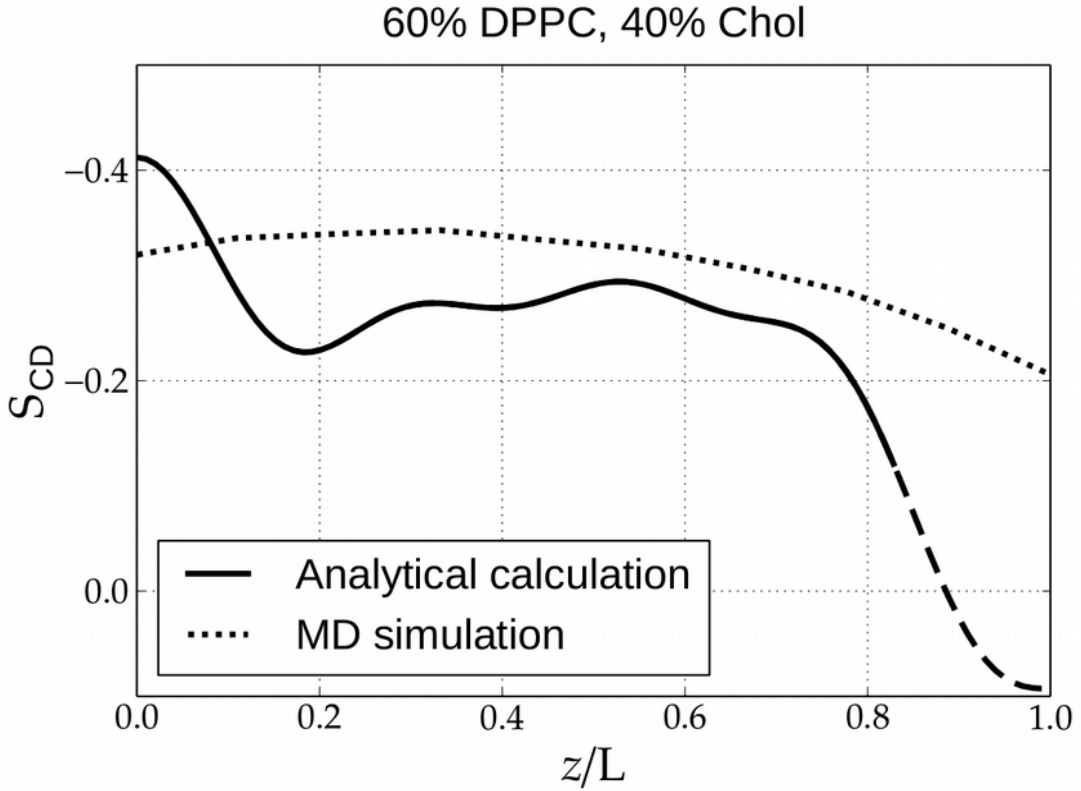

*Fig. 5: Solid line: calculated NMR order parameter. Dotted line: DPPC order parameter data from molecular dynamics simulation [18]. Both curves are plotted for temperature $T = 323$ K. With large $z/L$ approximation of small deviations of centers from the vertical in our model no longer works, and hence we plotted it with dashes.*

The result of analytical calculation resembles that of molecular dynamics simulation in the middle area of the monolayer, with two sufficient discrepancies in the borders: $S_{CD}(0) \approx -0.4$ value at which analytical curve starts, which is lower than the MD value, and the drop toward the midplane region, $S_{CD}(1)$.

The first discrepancy might be due to our model doesn't take into account the lateral oscillation of lipid as a whole, combined with a boundary condition [23], that chain should be parallel to the membrane normal near the membrane surface. We took into account gel phase tilt angle, which made it $\approx -0.4$ instead of $-0.5$. The value $S_{CD}(0)$ should be lowered by the cholesterol oscillation in the direction of membrane normal [32].

The second discrepancy is because small deviations of centers of the string from the vertical breaks in the midplane region.

## DMPC-cholesterol system

Supplement material for [33] contains data on NMR order parameter for, $S_{CD}$, for DMPC in a DMPC-cholesterol 1:1 molar mixture. DMPC has the same head group as DPPC and its tails are two methylene shorter than tails of DPPC. Hence, we modeled it with a shorter string, and we also increase incompressible area of the string describing DPPC lipid in order to obtain correct area per lipid in a pure DMPC membrane. The result of calculation is plotted on Fig. 6.

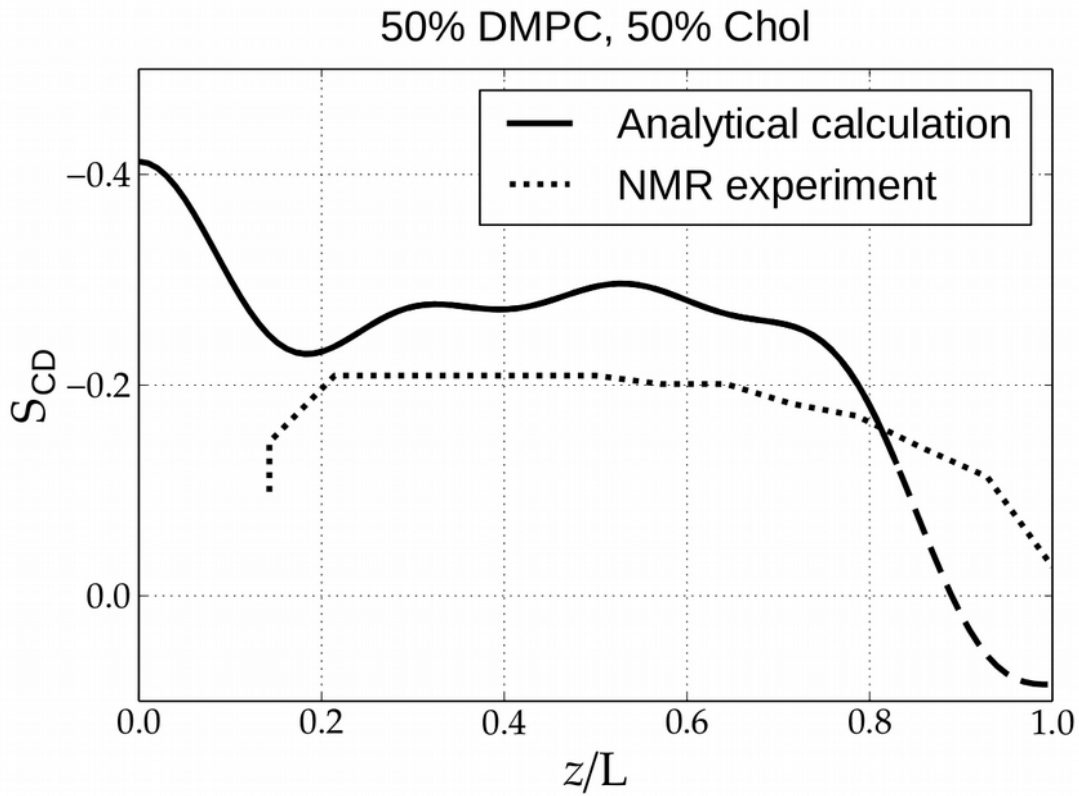

*Fig. 6: Solid line: calculated NMR order parameter. Dotted line: DMPC order parameter data from NMR experiment [33]. Both curves are plotted for T = 30 °C. With large z/L approximation of small deviations of centers from the vertical in our model no longer works, and hence we plotted it with dashes.*

Again the result of analytical calculation resembles that of NMR experiment, in the middle area of the monolayer, with sufficient differences in the borders: namely $S_{CD}(0)$ and $S_{CD}(1)$ values, which are caused by the limitations of our model.

## DOPC-DPPC system

Area per lipid in DOPC-DPPC systems is also non-linear [34].

DOPC and DPPC have same hydrophilic head and hence should be described with the same tension at the hydrophilic interface, $\gamma$ in our model. That means that $\gamma$ in DOPC-DPPC mixtures does not depend on concentration (see Eq. 10). And since $\gamma$ is the only parameter in our model that controls interaction of lipid with the membrane, lipids in DOPC-DPPC mixtures in our model wouldn't know that membrane is two component.

In [34] it is found that lipids in DOPC-DPPC mixtures form clusters. Our model in its current form merely describes a *cross-over* of membrane's tension at the hydrophilic interface. In order to describe clustorization it should be added with a direct lipid tails interaction.

## Conclusions

Flexible strings model is microscopic model of lipids in a membrane. It might be used to complement existing calculational techniques, such as theory of elasticity and computer simulations to get insight into the physics of lipid membranes.

In this paper we calculated the area per lipid in a DPPC-cholesterol bilayer membranes in a biologically relevant cholesterol concentrations. Few parameters of the model were tailored for an excellent agreement with the data available from molecular dynamics simulations (see Fig. 4).

We also found an inequality that must be satisfied by parameters of the model for a non-linear behavior of the area per molecule dependence on concentration of the components.

We then calculate NMR order parameter of DPPC for 60% DPPC – 40% cholesterol and DMPC for 50% DMPC – 50% cholesterol membranes, taking into account tilt of the DPPC and DMPC lipids. The results resemble data from molecular dynamics simulation and NMR experiments in the middle area of monolayer, with two discrepancies at the borders: i.e. for $S_{CD}(0)$ and $S_{CD}(1)$ values, the first one is because of the boundary condition we use [23], and the second one is due to small deviation of centers of the string from the vertical approximation doesn't work in a mid-plane region.

Finally, we note that our approach currently is limited to considering the homogeneous states of the membranes, e.g. without phase separation. Besides that, our simplified approach so far discerns different lipids merely by ascribing related with them different contributions to an effective tension $\gamma$ at the hydrophilic interface. These limitations will be lifted in the future.

## ACKNOWLEDGEMENTS

This work was supported in part by the Russian Foundation for Basic Research (project

no. 13 04 40327 H (KOMFI)). The authors are grateful to profs. R.G. Efremov, I.A. Boldyrev and colleagues for introduction into the problem.

## APPENDIX

Eq. 10 sets $\gamma$ of the mixture to be proportional to the molar concentration of the components. Nevertheless, setting $\gamma$ of the mixture to be proportional to the *area concentration* of the components

$$\gamma(c) = \gamma_C^0 \cdot \frac{A_C c}{c A_C + A_D(1-c)} + \gamma_D^0 \frac{A_D \cdot (1-c)}{c A_C + A_D(1-c)} \tag{32}$$

(here $c$ is a molar concentration of cholesterol) leads to a much more complex equations with the same result, as we show below.

In order to find area per lipid in a DPPC-cholesterol mixture with a $\gamma(c)$ of the form Eq. 32 one has to solve a system of equations for $A_C$ and $A_D$ for a given cholesterol concentration:

$$\begin{cases} \dfrac{\partial F_C}{\partial A} + \gamma(c) = 0 \\ \dfrac{\partial F_D}{\partial A} + \gamma(c) = 0 \end{cases} \tag{33}$$

where $\partial F_D / \partial A$ is given in Eq. 5, and $\partial F_C / \partial A$ is given in Eq. 8.

Substituting $A_C$ and $A_D$ in denominators of Eq. 32 with equilibrium values, one can find the solution of system 33 numerically. The obtained $\gamma(c)$ dependence is almost linear (see Fig. 7) and area per lipid curve is below the curve of ideal mixture (see Fig. 8), which is same we had with a much simpler Eq. 10.

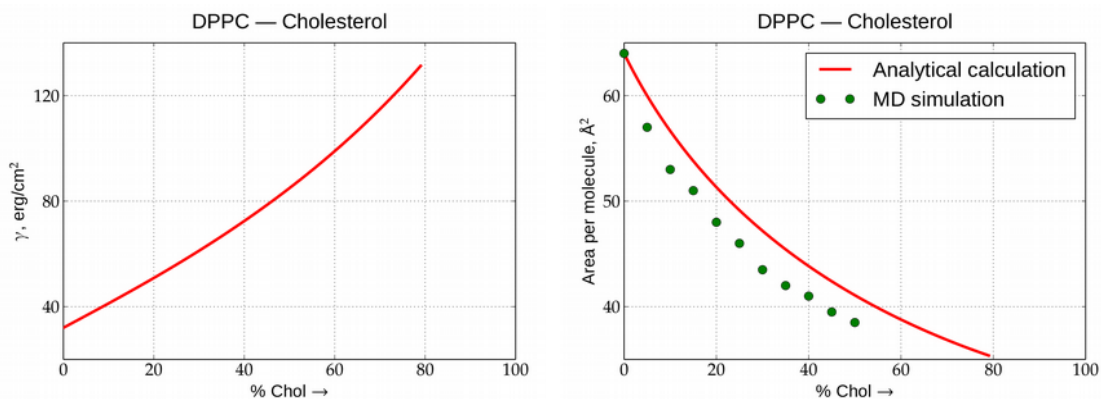

*Fig. 7: $\gamma(c)$ obtained by solving system 33 numerically.*

*Fig. 8: Area per lipid obtained by solving system 33 numerically.*

## REFERENCES

1. Bloom PD, Baikerikar K., Otaigbe JU, Sheares VV (2000) Development of novel polymer/quasicrystal composite materials. Mater Sci Eng A 294–296: 156–159.

doi:10.1016/S0921-5093(00)01230-2.

2. Brown DA, London E (2000) Structure and function of sphingolipid- and cholesterol-rich membrane rafts. J Biol Chem 275: 17221–17224. doi:10.1074/jbc.R000005200.

3. McMullen TP, McElhaney RN (1996) Physical studies of cholesterol-phospholipid interactions. Curr Opin Colloid Interface Sci 1: 83–90. Available: http://www.sciencedirect.com/science/article/pii/S1359029496800483.

4. Yeagle PL (1985) Cholesterol and the cell membrane. Biochim Biophys Acta 822: 267–287.

5. Alberts B (2008) Molecular biology of the cell. New York: Garland Science.

6. Needham D, McIntosh TJ, Evans E (1988) Thermomechanical and transition properties of dimyristoylphosphatidylcholine/cholesterol bilayers. Biochemistry (Mosc) 27: 4668–4673. Available: http://dx.doi.org/10.1021/bi00413a013.

7. Bloom M, Evans E, Mouritsen OG (1991) Physical properties of the fluid lipid-bilayer component of cell membranes: a perspective. Q Rev Biophys 24: 293–397.

8. Nakanishi M, Hirayama E, Kim J (2001) Characterisation of myogenic cell membrane: II. Dynamic changes in membrane lipids during the differentiation of mouse C2 myoblast cells. Cell Biol Int 25: 971–979. doi:10.1006/cbir.2001.0750.

9. Yip CM, Elton EA, Darabie AA, Morrison MR, McLaurin J (2001) Cholesterol, a modulator of membrane-associated Abeta-fibrillogenesis and neurotoxicity. J Mol Biol 311: 723–734. doi:10.1006/jmbi.2001.4881.

10. Chochina SV, Avdulov NA, Igbavboa U, Cleary JP, O'Hare EO, et al. (2001) Amyloid beta-peptide1-40 increases neuronal membrane fluidity: role of cholesterol and brain region. J Lipid Res 42: 1292–1297.

11. Cornelius F (2001) Modulation of Na,K-ATPase and Na-ATPase activity by phospholipids and cholesterol. I. Steady-state kinetics. Biochemistry (Mosc) 40: 8842–8851.

12. Simons K, Ikonen E (1997) Functional rafts in cell membranes. Nature 387: 569–572. doi:10.1038/42408.

13. Cantor RS (1999) Lipid Composition and the Lateral Pressure Profile in Bilayers. Biophys J 76: 2625–2639. Available: http://www.sciencedirect.com/science/article/pii/S0006349599774151.

14. Leathes JB (1925) Croonian Lectures ON THE RÔLE OF FATS IN VITAL PHENOMENA. The Lancet 205: 853–856. Available: http://www.sciencedirect.com/science/article/pii/S0140673601223101.

15. Róg T, Pasenkiewicz-Gierula M (2001) Cholesterol effects on the phospholipid condensation and packing in the bilayer: a molecular simulation study. FEBS Lett 502: 68–71.

16. Chiu SW, Jakobsson E, Mashl RJ, Scott HL (2002) Cholesterol-induced modifications in lipid bilayers: a simulation study. Biophys J 83: 1842–1853. Available: http://www.ncbi.nlm.nih.gov/pmc/articles/PMC1302277/.

17. McConnell HM, Vrljic M (2003) Liquid-liquid immiscibility in membranes. Annu Rev Biophys Biomol Struct 32: 469–492. doi:10.1146/annurev.biophys.32.110601.141704.

18. Hofsass C, Lindahl E, Edholm O (2003) Molecular Dynamics Simulations of Phospholipid Bilayers with Cholesterol. Biophys J 84: 2192–2206. Available: http://www.ncbi.nlm.nih.gov/pmc/articles/PMC1302786/.

19. Petrache HI, Tu K, Nagle JF (1999) Analysis of simulated NMR order parameters for lipid bilayer structure determination. Biophys J 76: 2479–2487. Available: http://www.ncbi.nlm.nih.gov/pmc/articles/PMC1300220/.

20. Tierney KJ, Block DE, Longo ML (2005) Elasticity and Phase Behavior of DPPC Membrane Modulated by Cholesterol, Ergosterol, and Ethanol. Biophys J 89: 2481–2493. Available: http://www.ncbi.nlm.nih.gov/pmc/articles/PMC1366747/.

21. Kessel A, Ben-Tal N, May S (2001) Interactions of cholesterol with lipid bilayers: the preferred configuration and fluctuations. Biophys J 81: 643–658. doi:10.1016/S0006-3495(01)75729-3.

22. Villalaín J (1996) Location of cholesterol in model membranes by magic-angle-sample-spinning NMR. Eur J Biochem FEBS 241: 586–593.

23. Mukhin SI, Baoukina S (2005) Analytical derivation of thermodynamic characteristics of lipid bilayer from a flexible string model. Phys Rev E 71: 061918. Available: http://link.aps.org/doi/10.1103/PhysRevE.71.061918.

24. Mukhin SI, Kheyfets BB (2010) Analytical Approach to Thermodynamics of Bolalipid Membranes. Phys Rev E 82: 051901. Available: http://link.aps.org/doi/10.1103/PhysRevE.82.051901.

25. Edholm O, Nagle JF (2005) Areas of Molecules in Membranes Consisting of Mixtures. Biophys J 89: 1827–1832. Available: http://www.sciencedirect.com/science/article/pii/S0006349505728277.

26. Mukhin SI, Kheyfets BB (2014) Pore formation phase diagrams for lipid membranes. JETP Lett 99: 358–362. Available: http://link.springer.com/article/10.1134/S0021364014060095.

27. Mukhin S.I., Drozdova A.A. . In preparation.

28. Landau L.D., Lifshitz E.M. (2008) Course of theoretical physics. Theory of Elasticity. Oxford [u.a.]: Butterworth-Heinemann.

29. Burkhardt TW (1995) Free energy of a semiflexible polymer confined along an axis. J Phys Math Gen 28: L629. Available: http://iopscience.iop.org/0305-4470/28/24/001.

30. Lindahl E, Edholm O (2000) Mesoscopic undulations and thickness fluctuations in lipid bilayers from molecular dynamics simulations. Biophys J 79: 426–433. Available: http://www.ncbi.nlm.nih.gov/pmc/articles/PMC1300946/.

31. Douliez J-P, Ferrarini A, Dufourc E-J (1998) On the relationship between C-C and C-D order parameters and its use for studying the conformation of lipid acyl chains in biomembranes. J Chem Phys 109: 2513–2518. Available: http://scitation.aip.org/content/aip/journal/jcp/109/6/10.1063/1.476823.

32. Gliss C, Randel O, Casalta H, Sackmann E, Zorn R, et al. (1999) Anisotropic motion of cholesterol in oriented DPPC bilayers studied by quasielastic neutron scattering: the liquid-ordered phase. Biophys J 77: 331–340. doi:10.1016/S0006-3495(99)76893-1.

33. Leftin A, Molugu TR, Job C, Beyer K, Brown MF (2014) Area per Lipid and Cholesterol Interactions in Membranes from Separated Local-Field 13C NMR Spectroscopy. Biophys J 107: 2274–2286. Available: http://www.cell.com/article/S0006349514007899/abstract.

34. Pyrkova DV, Tarasova NK, Pyrkov TV, Krylov NA, Efremov RG (2011) Atomic-scale lateral heterogeneity and dynamics of two-component lipid bilayers composed of saturated and unsaturated phosphatidylcholines. Soft Matter 7: 2569–2579. Available: http://pubs.rsc.org/en/content/articlelanding/2011/sm/c0sm00701c.